\documentclass[preprint]{iucr}
\journalcode{J}

\usepackage{color,xcolor}
\usepackage{xspace}
\RequirePackage{graphicx}
\usepackage{epstopdf}
\usepackage{mathtools, amsmath, amsthm, amssymb, amsfonts,bm}
\usepackage[12pt]{moresize}
\usepackage{multirow}
\usepackage{enumitem}
\usepackage{url}

\graphicspath{{./figures/}}

\definecolor{ykcolor}{HTML}{4d1063}

\renewcommand{\vec}[1]{\ensuremath\mathbf{#1}}
\newcommand{\be}{\begin{enumerate}[wide, labelwidth=!, labelindent=0pt,
        label=\textbf{\textcolor{blue}{\arabic*}.}]}
    \newcommand{\bei}{\begin{enumerate}}
        \newcommand{\ee}{\end{enumerate}}
    \newcounter{saveenumi}

\newcommand{\est}{$\sim$}

\newcommand{\dd}{\mathrm{d}}

\newcommand{\nba}[1]{}

\definecolor{dgreen}{HTML}{008000}

\newcommand{\qmax}{\ensuremath{Q_{\mathrm{max}}}\xspace}
\newcommand{\qmin}{\ensuremath{Q_{\mathrm{min}}}\xspace}

\newcommand{\pdfgui}{\textsc{PDFgui}\xspace}
\newcommand{\diffpy}{\textsc{diffpy}\xspace}
\newcommand{\pdfgetxthree}{\textsc{PDFgetX3}\xspace}

\newcommand{\pyfai}{\textsc{pyFAI}\xspace}

\newcommand{\saspdf}{\textsc{sasPDF}\xspace}
\newcommand{\pdfgets}{\textsc{PDFgetS3}\xspace}

\usepackage{stackengine}
\def\subrangle#1{\stackengine{5pt}{}{$\!\scriptstyle #1$}{U}{l}{F}{F}{L}}
\makeatletter
\let\save@rangle\rangle
\def\rangle{\save@rangle\@ifnextchar_{\expandafter\subrangle\@gobble}{}}
\makeatother

\makeatletter
\newcommand{\floatcaption}{%
    \ifx \@captype \@undefined \@latex@error {\noexpand \caption outside float}\@ehd \expandafter \@gobble \else \refstepcounter \@captype \expandafter \@firstofone \fi {\@dblarg {\@caption \@captype }}%
}%
\makeatother

\newcommand{\revisadd}[1]{#1}
\newcommand{\revisdel}[1]{}

\begin{document}

\title{sasPDF: pair distribution function analysis of nanoparticle assemblies from small-angle-scattering data}

\author[a]{Chia-Hao}{Liu}
\author[b]{Eric M.}{Janke}
\author[c]{Ruipen}{Li}
\author[d]{Pavol}{Juh\'{a}s}
\author[a,e,f]{Oleg}{Gang}
\author[b]{Dmitri V.}{Talapin}
\author[a,g]{Simon J. L.}{Billinge}

\aff[a]{Department of Applied Physics and Applied Mathematics, Columbia University, \city{New York}, NY~10027, \country{USA}}
\aff[b]{Department of Chemistry, University of Chicago, \city{Chicago}, IL~60637, \country{USA}}
\aff[c]{National Synchrotron Light Source II, Brookhaven National Laboratory, \city{Upton}, New York~11973, \country{USA}}
\aff[d]{Computational Science Initiative, Brookhaven National Laboratory, \city{Upton}, NY~11973, \country{USA}}
\aff[e]{Center for Functional Nanomaterials, Energy and Photon Sciences Directorate, Brookhaven National Laboratory, \city{Upton}, New York~11973, \country{USA}}
\aff[f]{Department of Chemical Engineering, Columbia University, \city{New York}, NY~10027, \country{USA}}
\aff[g]{Condensed Matter Physics and Materials Science Department, Brookhaven National Laboratory,\city{Upton}, New York~11973, \country{USA}}

\date{\today}

\maketitle

\section{Introduction}

With the advent of high degrees of control over nanoparticle synthesis~\cite{murraySynthesisCharacterizationNearly1993,hyeonSynthesisHighlyCrystalline2001,demellodonegaPhysicochemicalEvaluationHotInjection2005} attention is turning to assembling superlattices of them as metamaterials~\cite{bolesSelfAssemblyColloidalNanocrystals2016,choiExploitingColloidalNanocrystal2016} and applications of nanoparticle assemblies (NPA) based devices such as solar cells and field effect transistors have been demonstrated~\cite{talapinPbSeNanocrystalSolids2005,sargentSolarCellsPhotodetectors2008,talapinProspectsColloidalNanocrystals2010}.
It is crucial to study the structures of these NPAs if their properties are to be optimized.  For example, it has been shown that the mechanical~\cite{akcoraAnisotropicSelfassemblySpherical2009}, optical~\cite{youngUsingDNADesign2014}, electrical~\cite{vanmaekelberghElectronconductingQuantumDot2005a} and magnetic~\cite{sunMonodisperseFePtNanoparticles2000} properties can be further engineered by controlling the spatial arrangement of the constituents in the NPA.

Getting detailed quantitative structural information from NPAs, especially in 3D, is a challenging and largely unsolved problem.
Small angle scattering and electron microscopy (EM) have been the major techniques for studying the structure of NPAs~\cite{murraySynthesisCharacterizationMonodisperse2000d,talapinQuasicrystallineOrderSelfassembled2009}.
The technique of TEM yields high-resolution images of NPAs.
To obtain quantitative structural information it is necessary to either analyze the images manually~\cite{wangTransmissionElectronMicroscopy2000} or match observed images with patterns that are algorithmically generated from known structures~\cite{shevchenkoStructuralDiversityBinary2006}.
This approach can yield the structure types~\cite{zhuangControllableSynthesisCu2S2008} but does not typically result in the kind of quantitative 3D structural information we are used to obtaining for atomic structures of crystals, including accurate inter-particle vectors and distributions of inter-particle distances, or the range of structural coherence of the packing order.
It is desirable to explore scattering approaches that can yield that kind of information.

The technique of small-angle x-ray or neutron scattering (SAS) has been an important tool to study objects that have sizes from nano- to micrometer length-scales~\cite{turkevichLowAngleXRay1951,glatterNewMethodEvaluation1977,guinierXrayDiffractionCrystals1994,kochSmallangleScatteringView2003},
such as large nanocrystals~\cite{polteNucleationGrowthGold2010} and biological molecules~\cite{kochSmallangleScatteringView2003}, yielding information about the intrinsic shape, size distributions and scattering density of objects on these scales~\cite{glatterNewMethodEvaluation1977,beaucageApproximationsLeadingUnified1995,pedersenAnalysisSmallangleScattering1997,volkovUniquenessInitioShape2003,beaucageParticleSizeDistributions2004}.

When these nanoscale objects aggregate, correlation peaks appear in the SAS data resembling atomic-scale interference peaks (Diffuse scattering and Bragg peaks), but encoding information about particle packing rather than atomic packing~\cite{murraySynthesisCharacterizationMonodisperse2000d,nykypanchukDNAguidedCrystallizationColloidal2008a}.
Obtaining structural information about the NPAs from these correlation peaks appears to be a promising approach.
Although the recent developments in SAS modeling demonstrates the ability to account for phase, morphology and orientations of NPs in a lattice~\cite{yagerPeriodicLatticesArbitrary2014,luUnusualPackingSoftshelled2019}, fitting the SAS data with robust structural models to obtain quantitative information about the structure has barely been explored~\cite{macfarlaneNanoparticleSuperlatticeEngineering2011}

On the other hand, the atomic pair distribution function (PDF) analysis of x-ray and neutron powder diffraction has proven to be a powerful tool
for characterizing local order in materials, and for extracting quantitative structural information~\cite{proff;zk05,egami;b;utbp12,zobel;s15,keen;n15} when the
atoms are not long-range ordered, as is the case in nanoparticles.
Here we extend PDF analysis to handle correlation peaks in the small angle scattering data, allowing us to study the arrangement of particles in nanoparticle assemblies using the same quantitative modeling tools that are available for studying the atomic arrangements in nanoparticles themselves.
We describe the extension of the PDF equations in the small-angle scattering (SAS) regime and describe the data collection protocol for optimum data quality.
We also present the \pdfgets software package that can be readily used to extract the PDF from small-angle scattering data.
We then apply the \saspdf method to investigate structures of some representative NPA samples with different levels of structural order.

\section{Samples}

To test the method we obtained SAS data from the samples listed in Table~\ref{tab:sample_info}.
\begin{table*}
    \centering
    \floatcaption{Nanoparticle assemblies (NPA)
        considered in this study. Building block indicates the NP
        and surfactant linkers used to build the assemblies. $D$ is the
        particle diameter (one standard deviation in parentheses) estimated from
        TEM images and reported in the original publications listed in the Ref. column.
        Beamline is the x-ray
        beamline where the SAXS data were measured (see text for details).  PMA
        is Poly(methyl acrylate) and DDT is dodecanethiol.}
    \label{tab:sample_info}
    \vspace{1em}
        \begin{tabular}{ccccc}
            \hline
            Sample    & Building block     & $D$ (nm) & Beamline & Ref.\\
            \hline
            Au NPA   & DNA-capped Au NP & $11.4(1.0)$ & X21 & \cite{nykypanchukDNAguidedCrystallizationColloidal2008a}\\
            Cu$_2$S NPA & DDT-capped Cu$_2$S NP & $16.1(1.3)$   & 11-BM &\cite{hanSynthesisShapeTailoringCopper2008}\\
            SiO$_2$ NPA & PMA-capped SiO$_2$ NP & $14(4)$      & 11-BM & \cite{bilchakPolymerGraftedNanoparticleMembranes2017}\\
            \hline
        \end{tabular}
\end{table*}
Synthesis details of these NPA samples can be found in the references listed in the table.

\section{\saspdf method}

\begin{figure}
    \includegraphics[width=0.6\columnwidth]{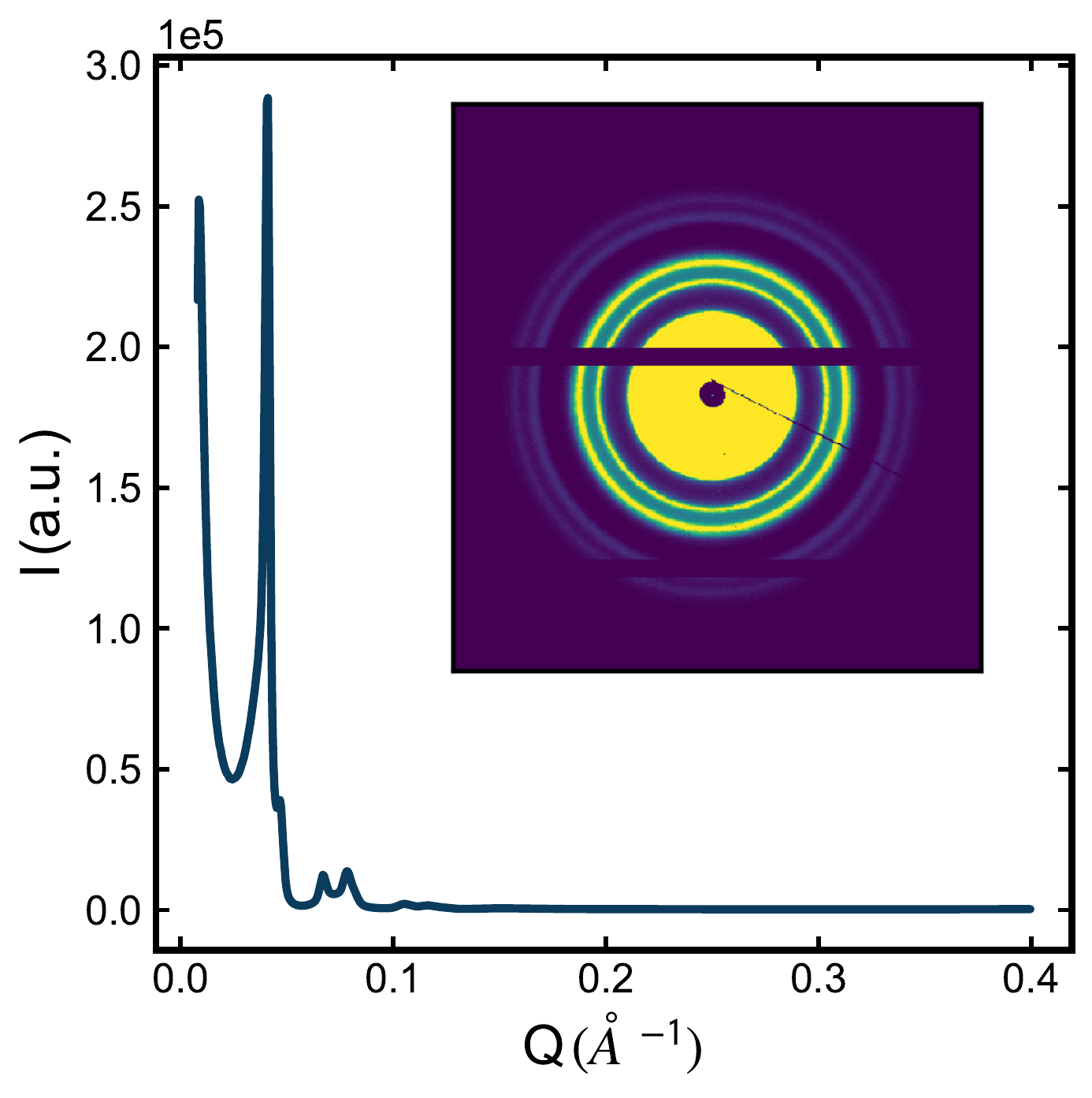}
    \label{fig:Iq_2Dimg_example}
    \caption{Example of the 1D diffraction pattern $I_m(Q)$ from the Cu$_2$S NPA sample. The data were collected with the spot exposure time and scan exposure time reported in the text. The inset shows the corresponding 2D diffraction image.
        The horizontal stripes in the image are from the dead zone between panels of the detector.
        The diagonal line is the beam-stop holder.}
\end{figure}

The data were collected using a standard SAXS setup at an x-ray synchrotron source, with a 2D area detector mounted perpendicular to the beam in transmission geometry.
Both the Cu$_2$S NPA and the SiO$_2$ NPA samples were measured at beamline 11-BM at the National Synchrotron Light Source-II (NSLS-II).
The Cu$_2$S NPA powders were sealed between two rectangular Kapton tapes with a circular deposited area of diameter about $3$~mm and thickness about $0.2$~mm.
The SiO$_2$ NPA formed a circular, free-standing stable film of diameter about $5$~mm and thickness about $1$~mm which was mounted perpendicular to the beam and no further sealing was carried out.
The scattering data of the Cu$_2$S NPA and SiO$_2$ NPA samples were collected with a Pilatus 2M (Dectris, Switzerland) detector with a sample-detector distance $2.02$~m using an x-ray wavelength of 0.918~\AA.
An example of the diffraction image from the Cu$_2$S NPA is shown in the inset of Fig.~\ref{fig:Iq_2Dimg_example}.
The scattering from these samples is isotropic as the sample consists of powders of randomly oriented NPA crystallites, and the 2D diffraction images can be reduced to a 1D diffraction pattern, $I_m(Q)$, by performing an azimuthal integration around rings of constant scattering angle on the detector.
This was done using \pyfai~\cite{kief;jac15}.
This requires a calibration of the experiment geometry described below, but the integrated 1D pattern from the 2D diffraction image is shown in Fig.~\ref{fig:Iq_2Dimg_example}.
The relative positions and intensities of sharp peaks in the $I_m(Q)$ originate from the Debye-Scherrer rings in the 2D image. 

We need to use a data acquisition strategy that mitigates effects of x-ray beam-damage to the sample.
The linkers that connect nanoparticles in the assemblies play a crucial role for the NPA structure formed but are susceptible to degradation in the intense x-ray beam that may result in changes in the NPA structure.
To describe the strategy we separate the concepts of a ``spot exposure time'' and the ``scan exposure time".
The latter is the total integrated exposure time to obtain a dataset with sufficient statistics.
The former is the length of time that any spot on the sample is exposed. The scan exposure then consists of multiple spot exposures, where the sample is translated after each spot exposure so that a fresh region of sample is exposed.
For ease of experimentation we would like to determine a spot exposure time
that is as long as possible whilst ensuring that the sample has not degraded significantly during that exposure.
We have found that the maximum safe spot exposure time depends on the nature of the NPA sample, as well as experimental conditions such as x-ray energy, flux and sample temperature.
It therefore requires a trial-and-error approach to determine it.
To choose the optimal spot exposure time we locate the beam on a fixed spot of the sample and take a sequence of short exposures, monitoring for significant changes in the intensity of the strongest correlation peak in $I_m(Q)$.
The spot exposure time determined this way for our experimental setup was 30~s for both Cu$_2$S NPA and SiO$_2$ NPA samples and the scan exposure time was 5~minutes (30~s, 10 spots) for the Cu$_2$S NPA sample and 10~minutes (30~s, 20 spots) for the SiO$_2$ NPA sample.

The scan exposure time is estimated based on an assessment of noise in the PDF given a desired $Q_{max}$, but it depends sensitively on the counting statistics in the high-$Q$ region of the diffraction pattern, which is easiest to assess by looking in the high-$Q$ region of the reduced structure function $F(Q)$.
For illustration purposes, the effect of scan exposure time on the $F(Q)$ (and the resulting PDF) is illustrated in  Fig.~\ref{fig:spot_exposure_example} of Appendix section.

For the calibration of the experimental geometry, such as sample-detector distance and detector tilting we use the calibration capability in the Python package \pyfai~\cite{kief;jac15}.
We first measured silver behenate (AgBh)~\cite{gillesSilverBehenatePowder1998} as a well characterized calibration sample.
The $d$-spacing of the calibration sample, the x-ray wavelength and the pixel dimensions of the detector are known, which allows the geometric parameters to be refined in \pyfai.
We found that selecting the strongest few rings (even just two or three work well) in the image allowed a stable refinement of the calibration parameters.

Finally, in this study we also consider legacy data from measurements carried out previously~\cite{nykypanchukDNAguidedCrystallizationColloidal2008a}.
The data of the Au NPA sample were collected at beamline X21 at the National Synchrotron Light Source (NSLS) from a sample loaded into a wax-sealed 1~mm diameter quartz capillary.
The scattering data were collected with a MarCCD (Rayonix, USA) area detector using an x-ray wavelength of 1.55~\AA.
Details of the measurements are reported in~\cite{nykypanchukDNAguidedCrystallizationColloidal2008a}.

The PDF, denoted $G(r)$, is a truncated sine Fourier transform of the reduced structure function $F(Q) = Q\left[(S(Q)-1)\right]$~\cite{egami;b;utbp12}
\begin{align}
\label{eq:FTofSQtoGr}
G(r) = \frac{2}{\pi}
\int_{\qmin}^{\qmax} F(Q)\sin(Qr) \: \dd Q.
\end{align}
Since $F(Q)$ can be easily computed once $S(Q)$ is available, we will first focus on describing the precise definition of $S(Q)$ and its relation to the measured diffraction pattern $I_m(Q)$. The measured intensity, $I_m(Q)$, depends on experimental details such as the flux, and beam size of the x-ray source, the data collection time and the sample density.
From the point of developing the \saspdf formalism, we will focus on the coherent scattering intensity $I_c(Q)$~\cite{egami;b;utbp12} which is obtained after correcting $I_m(Q)$ for the experimental factors as we describe below.

The coherent scattering intensity $I_c(Q)$ from a unit cell with $N_s$ atoms is~\cite{egami;b;utbp12,guini;b;xdic63}
\begin{equation}
\label{eq:Ic_def_0}
I_c(\vec{Q}) = \sum_{m=1}^{N_s}\sum_{n=1}^{N_s} f^*_m(\vec{Q}) f_n(\vec{Q}) \exp\left[i\vec{Q}\cdot(\vec{r}_m-\vec{r}_n)\right],
\end{equation}
where $\vec{Q}$ is the scattering vector, $f_m(\vec{Q})$ and $\vec{r}_m$ are the atomic form factor amplitude and position of $m$-th atom in the unit cell, respectively.

If the scattering from a sample is isotropic, for example, it is an untextured powder or a liquid with no anisotropy, the observed scattering intensity will depend only on the magnitude of $\vec{Q}$, $|\vec{Q}|=Q$ and not its direction in space.
The observed scattering intensity in this case will depend on the orientationally averaged $I_c(\vec{Q})$,
\begin{equation}
\label{eq:Ic_def_1}
I_c(Q) = \left\langle\sum_{m=1}^{N_s}\sum_{n=1}^{N_s} f^*_m(\vec{Q}) f_n(\vec{Q}) \exp\left[i\vec{Q}\cdot(\vec{r}_m-\vec{r}_n)\right]\right\rangle,
\end{equation}
where $\langle \cdot \rangle$ denotes the orientational average.

This formalism is readily extended to the case where the scattering objects are not atoms, but are some other finite-sized object, for example, nanoparticles.
In this case, the atomic form-factor would be replaced with the form-factor for the scattering objects in question.
The form factor $f(\vec{Q})$ for a generalized scatterer, with volume $V$ and its electron density as a function of position $\rho(\vec{r})$ is~\cite{guini;b;xdic63}
\begin{align}
\label{eq:ff_def}
f(\vec{Q}) = \int_V\left[\rho(\vec{r})-\rho_0\right]\exp\left(i\vec{Q}\cdot \vec{r}\right)\dd\vec{r},
\end{align}
where $\rho_0$ is the average electron density of the ambient environment of the scatterers.

In situations where there is only one type of scatterer we pull the form factors out of the sum\revisadd{~\cite{egami;b;utbp12}}. \revisadd{Further, if the form factors and the structure factors are separable} Eq.~\ref{eq:Ic_def_1} may be further simplified to
\begin{equation}
\label{eq:Ic_1_scatt}
I_c(Q)= N_s\left\langle f^2(Q)\right\rangle + \left\langle f(Q)\right\rangle^2\left\langle\sum_{m=1}^{N_s}\sum_{n\neq m}^{N_s} \exp\left[i\vec{Q}\cdot(\vec{r}_m-\vec{r}_n)\right]\right\rangle.
\end{equation}
\revisadd{This is valid for spherical or nearly spherical shaped particles, and may be more broadly true \cite{guini;b;sas55,kotlarchykAnalysisSmallAngle1983,liSmallAngleXray2016a}
    , though if the shape of the particle results in an orientation that depends on the packing, for example a long axis lies along a particular crystallographic direction in the nanoparticle assembly, this approximation will not be ideal and Eq.~\ref{eq:Ic_def_1} would have to be used.  We have not tested how badly the approximation breaks down in these circumstances.}

Following the Faber-Ziman formalism~\cite{faber;pm65},
\begin{align}
\label{eq:Sq_FaberZiman_0}
S(Q) &= \frac{I_c(Q)}{N_s\langle f(Q) \rangle^2} - \frac{\langle f^2(Q) \rangle - \langle f(Q) \rangle^2}{\langle f(Q) \rangle^2},
\end{align}
we plug in $\langle f^2(Q) \rangle = \langle f(Q) \rangle^2$ and Eq.~\ref{eq:Sq_FaberZiman_0} becomes
\begin{align}
\label{eq:Iq_Sq_simple_def_atomic}
I_c(Q) = N_s\langle f^2(Q) \rangle S(Q).
\end{align}
This expression is equivalent to representing the scatterers as points at the position of their scattering center, convoluted with their electron distributions. The resulting structure function, $S(Q)$, yields the arrangement of scatterers in the sample.
This expression is often expressed in the SAS literature as~\cite{guini;b;xdic63}
\begin{align}
\label{eq:Iq_Sq_simple_def_NPA}
S(Q) = \frac{I_c(Q)}{N_sP(Q)}.
\end{align}
Where $P(Q)$ is equivalent to $\langle f^2(Q) \rangle$~\cite{guini;b;xdic63}, the orientational average of the square of the form-factor.
We note that, as with the atomic PDF, the above analysis can be generalized to the cases of scattering from multiple types of scatterers~\cite{kotlarchykAnalysisSmallAngle1983,yagerPeriodicLatticesArbitrary2014,senesiSmallangleScatteringParticle2015a} and in the SAS case approximate corrections for asphericity of the electron density~\cite{jonesDNAnanoparticleSuperlatticesFormed2010a,rossNanoscaleFormDictates2015,zhangSelectiveTransformationsNanoparticle2015},
may be applied.

To determine $S(Q)$ we need to have $P(Q)$.
$P(Q)$ can be computed from a given electron density, or directly measured.
For the case of a NPA sample, the precise scattering properties of the NP ensemble in the sample, including any polydispersity or distribution of geometric shapes, are not always known, therefore it is best to measure the form factor directly, as described below.
In general we do not know $N_s$ and all of the experimental factors (for example, the incident flux, multiple scattering and so on).
The algorithm~\cite{billi;jpcm13} that is widely used for carrying out corrections for these effects in the atomic PDF literature~\cite{juhas;jac13} is also suitable for the SAS data.
It takes advantage of our knowledge of the asymptotic behavior of the $S(Q)$ function to obtain an {\it ad hoc} but robust estimation of $S(Q)$ from the measured $I_m(Q)$.
This is described in detail in~\cite{juhas;jac13}.
The resulting scale of the PDF is not well determined, but when fitting models to the data this is not a problem~\cite{peter;jac03}, and in practice it gives close to a correct scale for high quality measurements.
Here we show that we can take the same approach to obtain the PDF from the measured SAS data here.

In the test experiments we describe here, in each case the form factor of the nanoparticles was obtained from a measurement.
The NPs are suspended in solvent at a sufficient dilution to avoid significant inter-particle correlations.
\revisadd{If the NPs start aggregating in the solution, a bump appears in the scattering intensity, at the $Q$-value corresponding to the nearest-neighbor distance, and observation of such a bump may indicate a problem with the form factor sample.}
The SAS signal of the dilute NP solution is measured with good statistics over the same range of $Q$ as the measurement of the nanoparticle assemblies themselves, and ideally on the same instrument.
The signal of the solvent and its holder is also measured and then subtracted from the SAS signal of the dilute NP solution to obtain the correct particle form factor signal.
We emphasize that it is important to measure exactly the same batch of NPs to have an accurate form factor for the NPA sample considered.

A form factor measured with high statistics is crucial as the signal in $I_c(Q)$ is weak in the high-$Q$ region and noise from the $P(Q)$ measurement can be significant in this region.
Fig.~\ref{fig:ff_spot_exposure_example} shows the effect on $F(Q)$ (and the resulting PDF),  when processed using $P(Q)$ from different scan exposure times.
It is clear that the statistics of the form-factor measurement has a significant effect on the results.
In cases where any signal in $P(Q)$ does not change rapidly it may be smoothed to reduce the effects of limited statistics, at the cost of possibly introducing bias if the smoothing is not done ideally.
This will be particularly relevant when the nanoparticles are not monodisperse, as is somewhat common.

The experimental PDF $G(r)$ is then obtained via the Fourier transformation, Eq.~\ref{eq:FTofSQtoGr}.
The success of the \saspdf method depends heavily on the good statistics (high signal-to-noise ratio) throughout the entire diffraction pattern $I_c(Q)$ and the form factor $P(Q)$, as important information about the structure may reside in the high-$Q$ region where the signal intensity is weak.
It is recommended to use intense radiation sources such as synchrotrons.
A comparison in data quality from an in-house instrument and a synchrotron source is shown in Fig.~\ref{fig:inhouse_vs_synchrotron} of Appendix section.

\section{Software}

To facilitate the \saspdf method, we implemented a \pdfgets software program for extracting the sasPDF from experimental data.
Information about obtaining the software is on the \diffpy organization website (https://www.diffpy.org).
The software is currently supported in Python 2 (2.7) and Python 3 (3.4 and above).  It requires a license and is free for researchers conducting open academic research, but other uses require a paid license.

\begin{figure}
    \includegraphics[width=0.6\columnwidth]{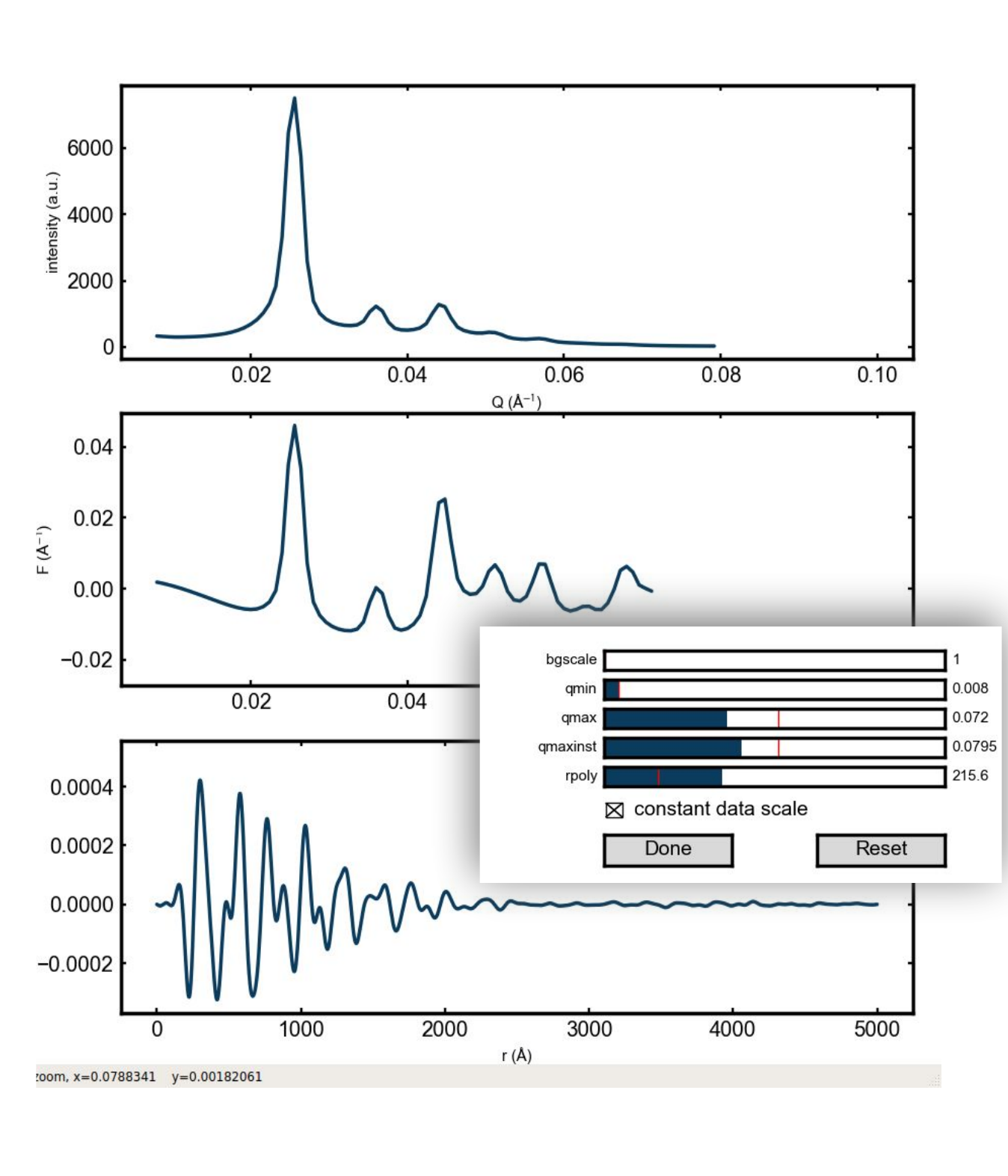}
    \label{fig:pdfgetsas_demo}
    \caption{Illustration of the interactive interface for tuning the process parameters in the \pdfgets program.}
\end{figure}

The \pdfgets program takes in a measured diffraction pattern $I_m(Q)$ and a form factor, $P(Q)$, as the inputs and applies a series of operations such as subtraction of experimental effects and form factor normalization and outputs the PDF, $G(r)$.
If the square of the orientationally averaged form-factor $\langle f(Q) \rangle^2$ is available, both $P(Q)$ and $\langle f(Q) \rangle^2$ can be specified in the program, and the $S(Q)$ will be computed based on Eq.~\ref{eq:Sq_FaberZiman_0} which accounts for the anisotropy of scatterers in the material.
Processing parameters used in \pdfgets operations, such as the form-factor file, the $Q$-range of the Fourier transformation on $F(Q)$ and the $r$-grid of the output $G(r)$, can be set in a configuration file in the same way detailed in~\cite{juhas;jac13}.
Similar to \pdfgetxthree, an interactive window for tuning these processing parameters, is also available in \pdfgets.
An illustration of such interactive interface is shown in the Fig.~\ref{fig:pdfgetsas_demo}.
Sliders for each processing parameter allow the user to inspect the effect on the output PDF data immediately.

Once the optimal processing parameters are determined based on the quality of the PDF, those parameter values will be stored as part of the metadata in the output $G(r)$ file.
The final values of \qmin and \qmax should be used when calculating PDF from a structure model, as these parameters contribute to the ripples in the PDF~\cite{peter;jac03}.
Full details on how to use the program is available on the \diffpy organization website.

\section{PDF method}
The PDF gives the scaled probability of finding two scatterers in a material a distance $r$ apart~\cite{egami;b;utbp12}.
For a macroscopic object with $N$ scatterers, the atomic pair density, $\rho(r)$, and $G(r)$ can be calculated from a known structure model using
\begin{equation}
\label{eq:def_rhor}
\rho(r) = \frac{1}{4 \pi r^{2} N}
\sum_{m}\sum_{n \neq m}
\frac{f_m(Q)f^*_n(Q)}{\langle f(Q) \rangle^2_{s.a.}}
\delta (r - r_{mn}),
\end{equation}
and
\begin{equation}\label{eq:Grfromrhor}
G(r) = 4 \pi r \left[ \rho(r) - \rho_{0} \right].
\end{equation}
Here, $\rho_{0}$ is the number density of scatters in the object.
$f_{m}(Q) = \langle f_m(\vec{Q}) \rangle $ is the orientationally averaged form-factor of the $m$-th scatterer.
$\langle f(Q) \rangle_{s.a.} = \sum_{m=1}^{N} (\frac{N_m}{N})f_m(Q)$ denotes the sample average of $f(Q)$ over all scatterers in the material, where $N_m$ is the number of scatterers that are of the same kind as the $m$-th scatter.
Finally, $r_{mn}$ is the distance between the $m$-th and $n$-th scatterer.
We use Eq.~\ref{eq:Grfromrhor} to fit the PDF generated from a structure model to a PDF determined from experiment.

PDF modeling, where it is carried out, is performed by adjusting parameters of the structure model, such as the lattice constants, positions of scatterers and particle displacement parameters (PDPs), to maximize the agreement between the theoretical and an experimental PDF.
In practice, the delta functions in Eq.~\ref{eq:Grfromrhor} are Gaussian-broadened to account for thermal motion of the scatterers and the equation is modified with a damping factor to account for instrument resolution effects.
The modeling of \saspdf can be done seamlessly with tools developed in the atomic PDF field, with parameter values scaled accordingly.
We outline the modeling procedure using \pdfgui~\cite{farro;jpcm07}, which is widely used to model the atomic PDF.
In \pdfgui, the nanoparticle arrangements can simply be treated analogously as atomic structures, with a unit cell and fractional coordinates, but the lattice constants reflect the size of the NPA, usually ranged from 10~nm = 100~\AA~to 100~nm = 1000~\AA.
The atomic displacement parameters (ADPs) defined in \pdfgui can be directly mapped to the particle displacement parameters (PDPs) in the \saspdf case and, empirically, we find the PDP values are roughly four to five orders of magnitude larger than the values of its counterpart on the atomic scale, therefore starting values of 500~\AA$^2$ are reasonable.
These will be adjusted to the best-fit values during the refinement.
\revisdel{The PDF peak intensity depends on the scattering length of relevant particle, which in the case of x-rays scattering  off atoms, is the atomic number of the atom.
For the \saspdf case we do not know explicitly how to scale the scattering strength of the particles, but for systems with a single scatterer, this constitutes an arbitrary scale factor that we neglect.}

The measured \saspdf signal falls off with increasing $r$.
The damping may originate from various factors, for example, the instrumental $Q$-space resolution~\cite{egami;b;utbp12} and finite range of order in the superlattice assembly.
In \pdfgui there is a a Gaussian damping function $B(r)$,
\begin{equation}
\label{eq:qdamp_def}
B(r) =
\exp\left[-\frac{\left(rQ_{damp}\right)^2}{2}\right].
\end{equation}
We define a $r_{damp}$ parameter
\begin{equation}
\label{eq:rdamp_def}
r_{damp} = \frac{1}{Q_{damp}},
\end{equation}
which is the distance where 
about one third of the \saspdf signal disappears completely.
It is also possible to generalize the modeling process to the case of a customized damping function and non-crystallographic structure with Diffpy-CMI~\cite{juhas;aca15}, which is a highly flexible PDF modeling program.
In the following section, we use \pdfgui for modeling data from more ordered samples (Au NPA and Cu$_2$S NPA) and Diffpy-CMI for modeling data from a disordered sample (SiO$_2$ NPA).

\section{Application to representative structures}
\begin{figure}
    \includegraphics[width=0.6\columnwidth]{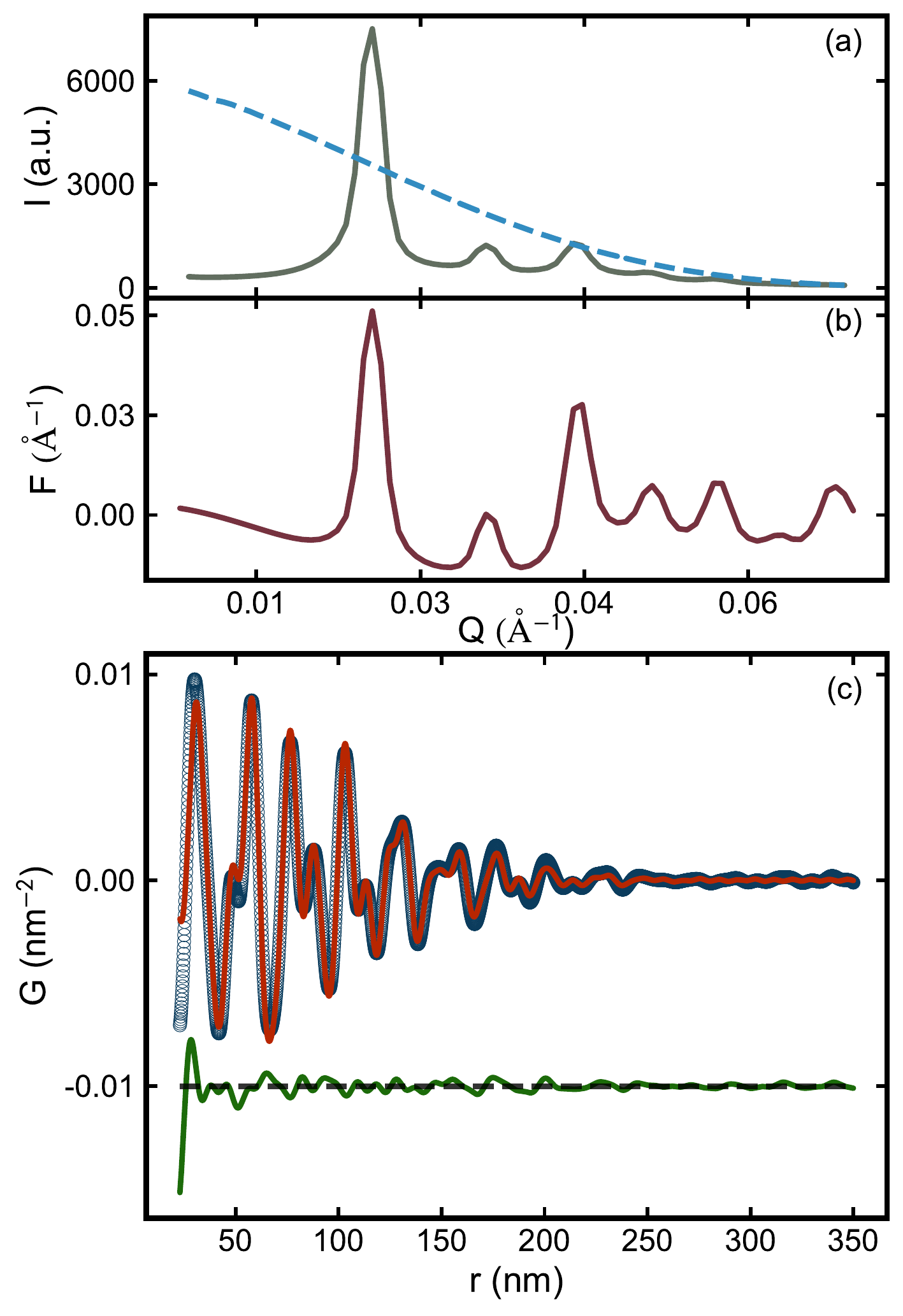}
    \label{fig:AuDNA_IqFqGr}
    \caption{Measured (a) scattering intensity $I_m(Q)$ (grey) and form factor $P(Q)$ (blue), (b) reduced total structure function $F(Q)$ (red) and (c) PDF (open circle) of Au NPA. In (c), the PDF calculated from body-center cubic (bcc) model is shown in red and the difference between the measured PDF and the bcc model is plotted in green with an offset.}
\end{figure}

To illustrate the \saspdf method we have applied it to some representative nanoparticle assemblies from the literature~\cite{nykypanchukDNAguidedCrystallizationColloidal2008a,hanSynthesisShapeTailoringCopper2008,bilchakPolymerGraftedNanoparticleMembranes2017}.
The first example is from DNA templated gold nanoparticle superlattices, originally reported in~\cite{nykypanchukDNAguidedCrystallizationColloidal2008a}.
The measured intensity, $I_m(Q)$, the reduced total structure function
$F(Q) = Q[S(Q)-1]$, and the PDF $G(r)$ are shown in Fig.~\ref{fig:AuDNA_IqFqGr}(a), (b) and (c), respectively.
It is clear that the data corrections and normalizations to get $F(Q)$ result in a more prominent signal in the high-$Q$ regime of the scattering data, and a highly structured PDF after the Fourier transform (Fig.~\ref{fig:AuDNA_IqFqGr}(c)).

The PDF signal dies off around $350$~nm, which puts a lower bound on the size of the NPA.
The first peak in the PDF is located at $30.07$~nm which corresponds to the nearest inter-particle distance in the assembly.
This distance is expected because the shortest inter-particle distance can be approximated as the average size of Au NPs ($11.4$~nm) plus the average surface-to-surface distance ($d_{ss}$) between nearest neighbor NPs ($18$~nm)~\cite{nykypanchukDNAguidedCrystallizationColloidal2008a}.
Peaks beyond the nearest neighbor give an indication of characteristic inter-particle distances in the assembly and codify the 3D arrangement of the nanoparticles in space.

A semi-quantitative interpretation of conventional powder diffraction data suggested the Au NPA forms a body-centered cubic (bcc)
structure~\cite{nykypanchukDNAguidedCrystallizationColloidal2008a}.
We therefore test the bcc model against the measured PDF.
The fit is shown in Fig.~\ref{fig:AuDNA_IqFqGr}(c) and the refined parameters are
reproduced in Table~\ref{tab:NPA_refined_param}.
\begin{table*}
    \centering
    \floatcaption{
        Refined parameters for NPA samples. Model column specifies the structural model used to fit
        the measured PDF. $a$ is the lattice constant of the unit cell,
        PDP stands for particle displacement parameters, which is an indication
        of the uncertainty in position of the nanoparticles. $r_{damp}$ is
        the standard deviation of the Gaussian damping function defined in
        Eq.~\ref{eq:rdamp_def}. Scale is a constant factor being multiplied to
        the calculated PDF.
        $R_w$ is the residual-function, commonly used as a measure for the goodness of fit.}
    \label{tab:NPA_refined_param}
    \vspace{1em}
    \begin{tabular}{ccc}
        & Au NPA & Cu$_2$S NPA \\
        \hline
        Model & bcc    & fcc      \\
        $a$~(nm)     & 34.73   & 26.55     \\
        PDPs~(nm$^2$)   & 4.78   & 0.253
        \\
        $r_{damp}$~(nm) & 83.3   & 61.4     \\
        Scale & 0.537  & 0.361    \\
        $R_w$ & 0.172  & 0.221
    \end{tabular}
\end{table*}
The agreement between the bcc model and the measured data is good.
We refine a lattice parameter that is \est3~\% smaller than the
value reported from the semi-quantitative analysis.
Additionally, the PDF gives information about the disorder in the system in the form of the crystallite size (\est350~nm) and the particle displacement parameter (PDP), the nanoparticle assembly version of the atomic displacement parameter (ADP)
in atomic systems.
The PDF derived crystallite size is drastically smaller than the value (\est500~nm) estimated from the FWHM of the first correlation peak~\cite{nykypanchukDNAguidedCrystallizationColloidal2008a} and
it is clear by visual inspection of the PDF that the \est500~nm
value is an overestimate.
These results suggest that even in the case where it is straightforward to infer the geometry of underlying assembly using qualitative and semi-quantitative means there is an advantage to carrying out a more fully quantitative \saspdf\ analysis.

\begin{figure}
    \includegraphics[width=0.6\columnwidth]{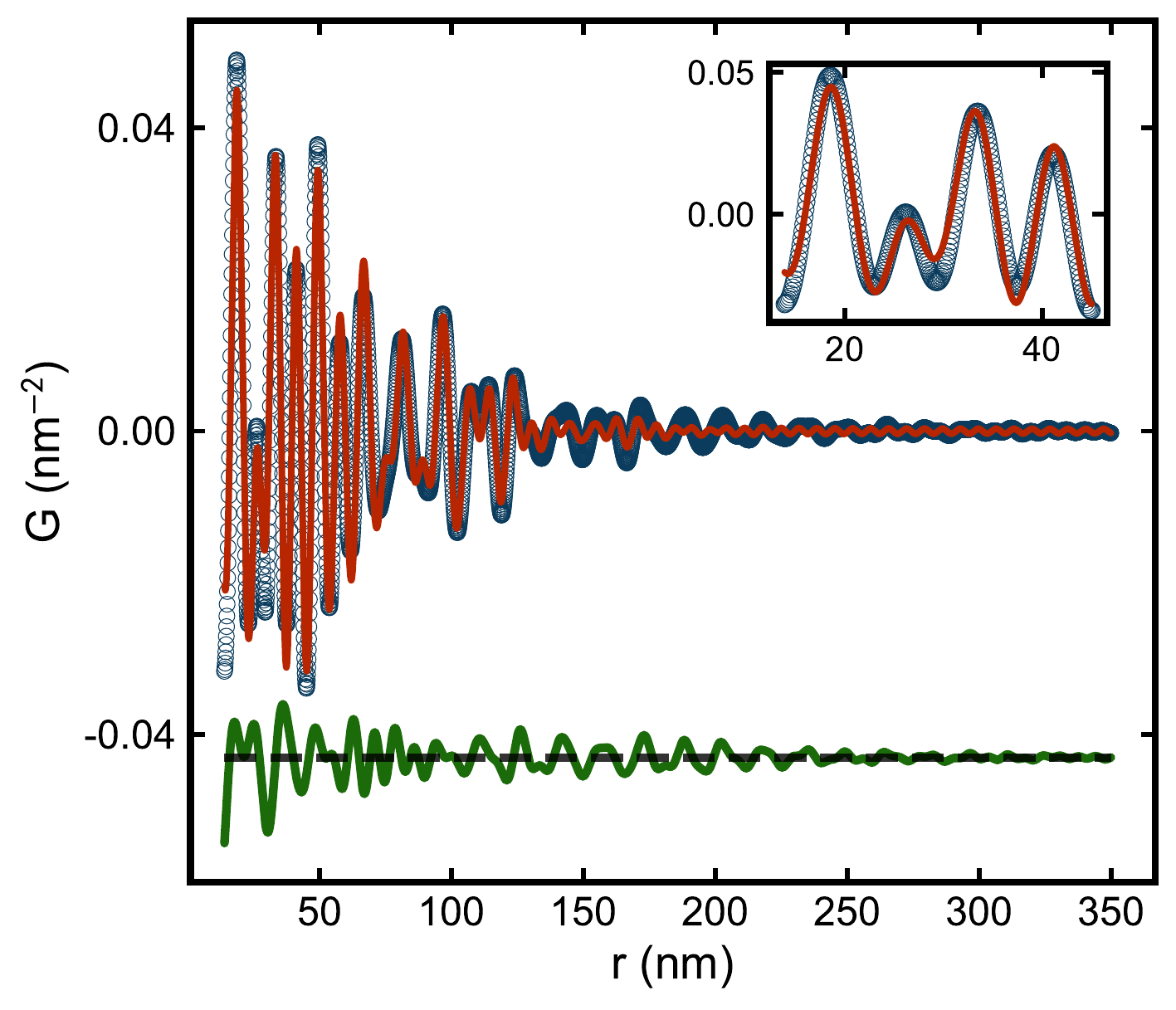}
    \label{fig:Cu2S_ordered_fcc}
    \caption{Measured PDF (open circle) of a Cu$_2$S NPA sample with the best fit PDF from the fcc model (red line). The Difference curve between the data and model is plotted with an offset in green.
        The inset shows the region of the first four nearest neighbor peaks of the PDF along with the best-fit fcc model.}
\end{figure}

Next we consider a dataset from a dodecanethiol (DDT)-capped Cu$_2$S NPA~\cite{hanSynthesisShapeTailoringCopper2008}.
In this case the form factor is measured on an in-house Cu K$_\alpha$ instrument.
This was necessary in the current case because the instability of the nanoparticles in suspension prevented a good measurement to be made at the synchrotron.
As a result the form factor measurement was somewhat noisy (Fig.~\ref{fig:smoothed_Cu2S_form_factor}(a), blue curve) and we elected to smooth it by applying a Savitzky-Golay filter~\cite{orfanidisIntroductionSignalProcessing1996}.
The smoothing parameters of window size and polynomial order were selected as 13 and 2, respectively, based on a trial and error approach optimized to result in a good smoothing without changing the shape of the signal. The smoothed curve is shown in Fig.~\ref{fig:smoothed_Cu2S_form_factor}(a).
It is worth noting that in general, a smoothing process may start failing when the signal-to-ratio in the data is below a certain threshold, and so good starting data is always desirable.
A conventional semi-quantitative analysis on diffraction data from the sample collected on an in-house Cu K$_\alpha$ instrument is shown in Fig.~\ref{fig:Cu2S_inhouse_Iq_indexing}.
It suggests the NPA forms a face-centered cubic (fcc) structure with an inter-particle distance of $18.8$~nm.
The SAS PDF obtained from the same NPA sample is shown in Fig.~\ref{fig:Cu2S_ordered_fcc}.
It clearly shows that peaks die out at around $300$~nm, which again signifies the crystallite size of the assembly.
The first peak of the measured PDF is at $18.5$~nm, corresponding to the inter-particle distance in the NPA.
This value is about $1.6$~\% smaller than the value estimated from the semiquantitative analysis.

The best-fit PDF of a close-packed face-centered cubic (fcc) structural model is shown in red in the figure and refined structural parameters are presented in Table~\ref{tab:NPA_refined_param}.
The fcc model yields a rather good agreement with the measured PDF of Cu$_2$S NPA in the short-range (up to \est130~nm).
Interestingly, the refined lattice parameter of this cubic model is $26.55$~nm, from which we can calculate an average inter-particle spacing of $18.78$~nm, which is much closer to the value estimated from the in-house data than directly extracting the position of the first peak in the PDF.
The first peak in the PDF calculated from the model lines up with that from the data at $18.5$~nm, which means that the position of the peak, as extracted from the peak maximum, underestimates the actual inter-particle distance by $\sim 1.5$\%, which may be due to the sloping background in the $G(r)$ function~\cite{egami;b;utbp12}.
Quantitative modeling is always preferred for obtaining the most precise determination of inter-particle distance.

The region of the first four nearest-neighbor peaks in the PDF, together with the fit, is shown in the inset to Fig.~\ref{fig:Cu2S_ordered_fcc}.
A close investigation of this region shows subtle shifts in peak positions between the measured PDF and the refined fcc model.
At around $26$~nm (second peak), the peak from refined model is shifted to higher-$r$ compared to the measured data, while at around $33$~nm (third peak), the relative shift in peak position is towards the low-$r$ direction.
These discrepancies suggest the NPA structure is more complicated than a simple fcc structure and may reflect the presence of internal twined defects, for example~\cite{baner;aca19}.
Furthermore, it is clear that signal persists in the measured PDF in the high-$r$ region that is not captured by the single-phase damped fcc model.
There is clearly more to learn about the structure of the NPA by finding improved structural models and fitting them to the PDF, though this is beyond the scope of the current paper.

It is worth noting that the refined PDP value of DDT-capped Cu$_2$S NPA is significantly smaller than that of the DNA-templated Au NPA described above.
A small PDP means the positional disorder of the NPs is small which would be expected with shorter, more rigid, linkers between the particles.
The inter-particle distance ($18.8$~nm) can be decomposed into the sum of the average particle diameter ($16.1$~nm) and the particle-surface to particle-surface distance $d_{ss}=2.7$~nm.
Based on the chemistry the linker would have length $1.7$~nm in the fully stretched out state, which would result in a maximal $d_{ss} =  3.4$~nm if the linkers were stretched out and oriented radially.
Half the observed surface-surface distance, $d_{ss}/2 = 1.4$~nm.
This result is reasonable, suggesting the linkers are either not straight, or not radial, or possibly partially interleaved.
Nonetheless, this shorter linker would be expected to be more rigid and therefore consistent with our observation of a smaller PDP value from the \saspdf analysis.

Finally we consider a dataset from a \revisadd{more disordered system}, poly(methyl acrylate) (PMA) capped SiO$_2$ NPA sample.
The molecular weight and density of the capping polymers can be varied and in the current sample was 0.47~chains/nm$^2$ and 132~kDa, respectively.
Studies had suggested that similar NPA samples exhibit no structural order, based on an empirical metric using the height of the first peak in the measured $S(Q)$~\cite{bilchakPolymerGraftedNanoparticleMembranes2017}.
Here we apply the \saspdf method to obtain a more \revisdel{quantitative}\revisadd{complete} understanding of the structure of this NPA.

\begin{figure}
    \includegraphics[width=0.6\columnwidth]{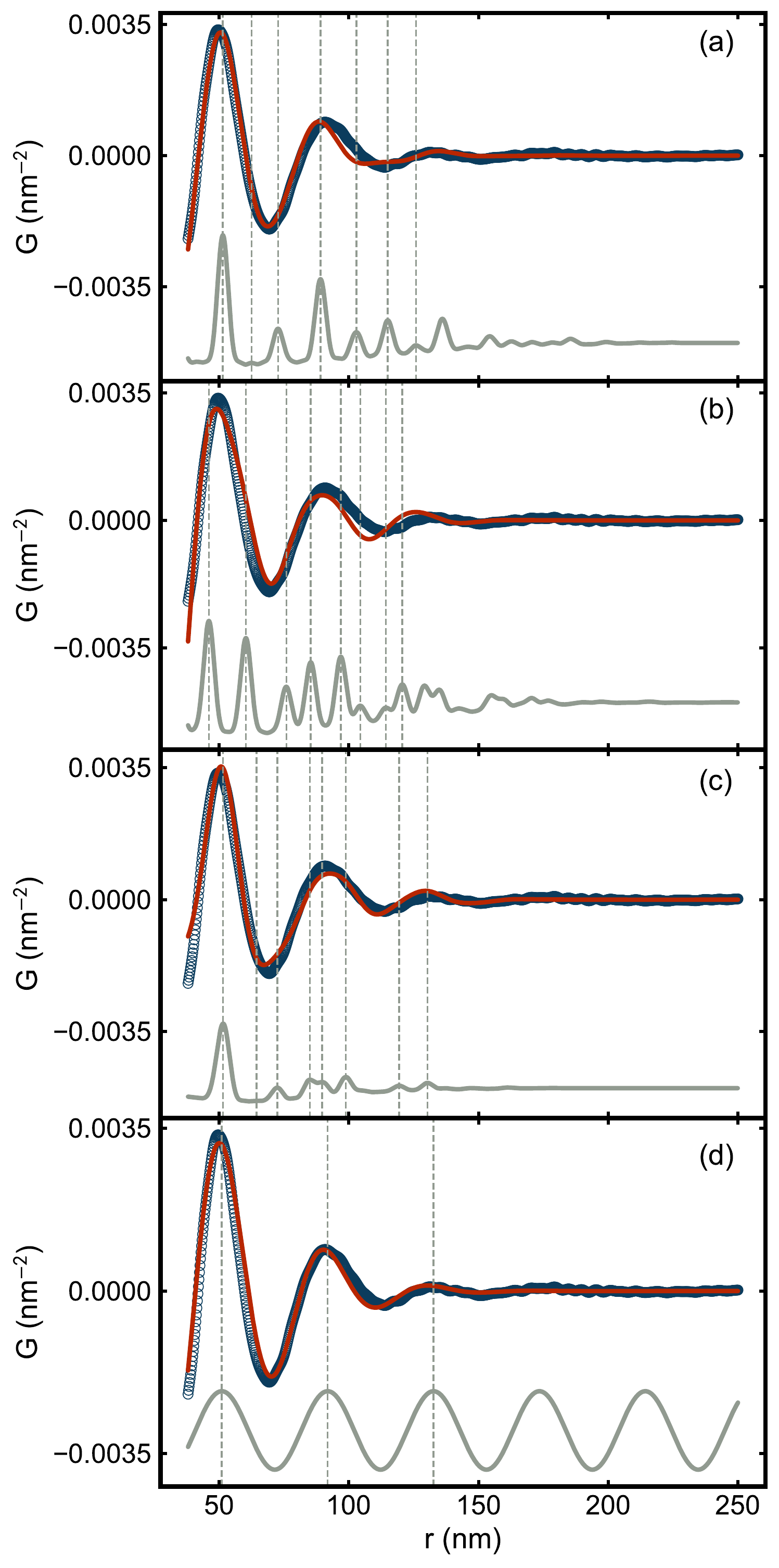}
    \label{fig:dispute_closepacked}
    \caption{Measured PDF (open circle) of the SiO$_2$ NPA sample and calculated PDFs (solid lines) from (a) fcc, (b) hcp, (c) icosahedral (d) damped sine-wave models. In each panel, the line in red is the PDF calculated from the corresponding model with optimum parameters. From (a) to (c), the line in grey is the PDF calculated from the same model but with small PDPs. In (d), the line in grey is the PDF calculated from the undamped sine-wave model. Dashed lines indicate maxima of the sharper PDFs in each panel.}
\end{figure}

To start we want to verify whether there is any evidence for close-packing of the NPs so we start with face-centered cubic (fcc), hexagonal close-packed (hcp) and icosahedral models~\cite{bausModernTheoryCrystallization1983} to see if any good agreement between the structural model and the data can be achieved.
The results are shown in Fig.~\ref{fig:dispute_closepacked}.
%
A highly broadened version of the fcc model yields a reasonable agreement with the data (Fig.~\ref{fig:dispute_closepacked}(a)).
The PDF of the fcc structure is shown in grey in the figure, and after significant broadening it yields the red curve.
\revisdel{However, closer inspection suggests that the fcc model is placing intensity at inter-particle distances where there is no intensity in the measured data, for example, at the positions of the
third and the fifth peak in the grey curve (the peak positions are indicated by vertical dashed lines).}
Other close-packed cluster models, hcp and icosahedral, were also \revisdel{similarly inadequate}\revisadd{tried} (Fig.~\ref{fig:dispute_closepacked}(b) and (c)).
Finally, a damped-sine wave model that is appropriate for highly disordered systems where on average the packing around a central atom is completely isotropic was tried (Fig.~\ref{fig:dispute_closepacked}(d)) \cite{cargillStructureMetallicAlloy1975,konnertComputationRadialDistribution1973,doann;acsn14}.
\revisdel{In particular, all three models fail in the region between 70~nm to 120~nm.
Since the close-packed models are failing we seek a different model for the local packing in the NPA.}\revisadd{The fcc and hcp structures would be expected for close-packed hard spheres.  Interestingly, in the current case, the very simple damped sine-wave model with only 3 parameters yields a more satisfactory fit to the data than the more complicated (with 4-5 parameter) close-packed models, suggesting that we have soft-sphere like packing in the current case.  This finding is being explored in more detail in another publication~\cite{liu;nm20}.}
This example shows how inter-particle packing may be examined using \saspdf method even when there is not long or intermediate range order.

\section{Acknowledgements}
This work was supported by the National Science Foundation DMREF program under CBET-16929502.
X-ray measurements conducted at beamline 11-BM of the National Synchrotron Light Source II, a U.S. Department of Energy (DOE) Office of Science user facility operated for the DOE Office of Science by Brookhaven National Laboratory was supported under Contract No. DE-SC0012704. Data collected at beamline X17A of the National Synchrotron Light Source which was supported by the DOE-BES under contract No. DE-AC02-98CH10886.
P. J. was supported by the New York State BNL Big Data Science Capital Project under the U.S. DOE Contract No. DE-SC0012704.
O.G. was supported by the US Department of Energy, Office of Basic Energy Sciences under the grant No. DE-SC0008772.
E.M.J. and D.V.T. were supported by the US Department of Energy, Office of Basic Energy Sciences (Grant No. DE-SC0019375), and by NSF under award number CHE-1905290


\appendix
\setcounter{figure}{0}
\setcounter{equation}{0}
\setcounter{table}{0}
\makeatletter
\renewcommand\thefigure{\thesection\arabic{figure}}
\renewcommand{\theequation}{S\arabic{equation}}
\renewcommand{\thetable}{S\arabic{table}}
\makeatletter

\section{Illustration of of data acquisition strategy}
\begin{figure}
    \includegraphics[width=0.6\columnwidth]{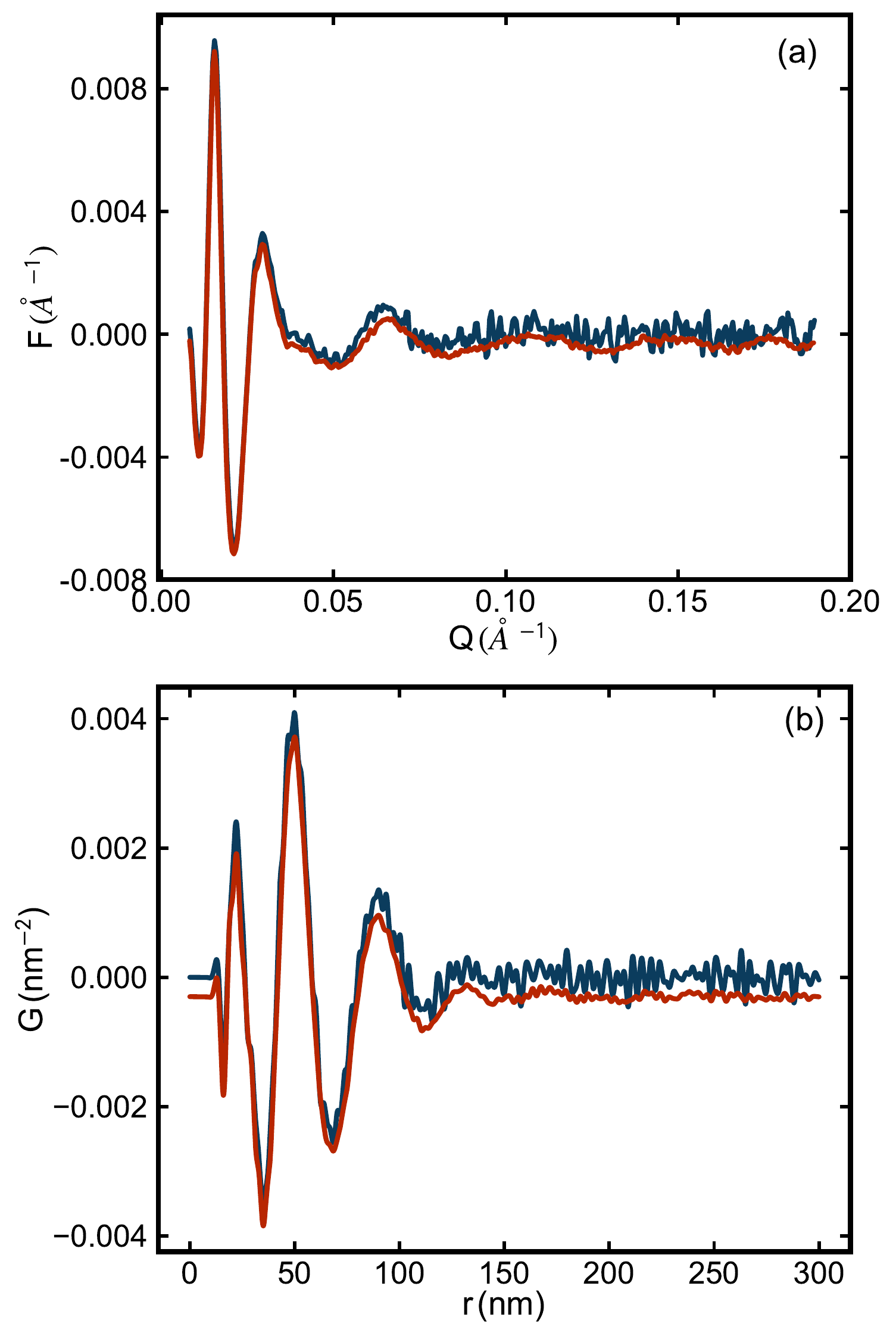}
    \label{fig:spot_exposure_example}
    \caption{(a) Reduced structure functions $F(Q)$ and (b) PDFs $G(r)$ of the SiO$_2$ NPA sample with different scan exposure times.
    Blue is from data with 1~s scan exposure time and red is from data with 30~s scan exposure time. In both panels, data are plotted with a small offset for ease of viewing. In both cases the form factor was measured with an scan exposure time of 600~s.}
\end{figure}

\begin{figure}
    \includegraphics[width=0.6\columnwidth]{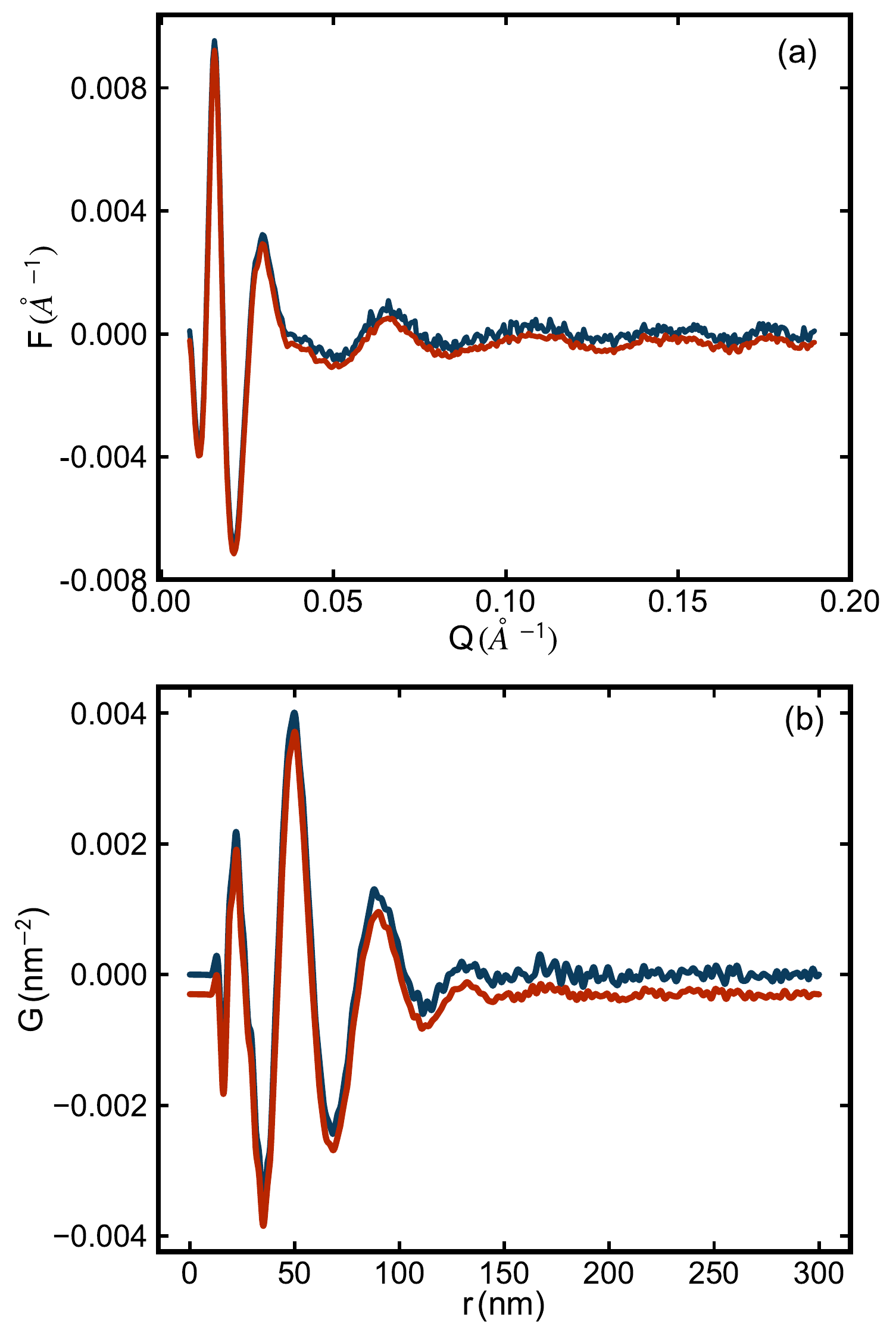}
    \label{fig:ff_spot_exposure_example}
    \caption{(a) Reduced structure functions $F(Q)$ and (b) PDFs $G(r)$ of the SiO$_2$ NPA sample processed with form factor $P(Q)$ from different scan exposure times.
    Blue is made with a form-factor measured for 30~s and red is with a form factor collected for 600~s. In both cases the scan exposure time for the NPA sample was 600~s. In both panels, data are plotted with a small offset for ease of viewing.}
\end{figure}

\begin{figure}
    \includegraphics[width=0.6\columnwidth]{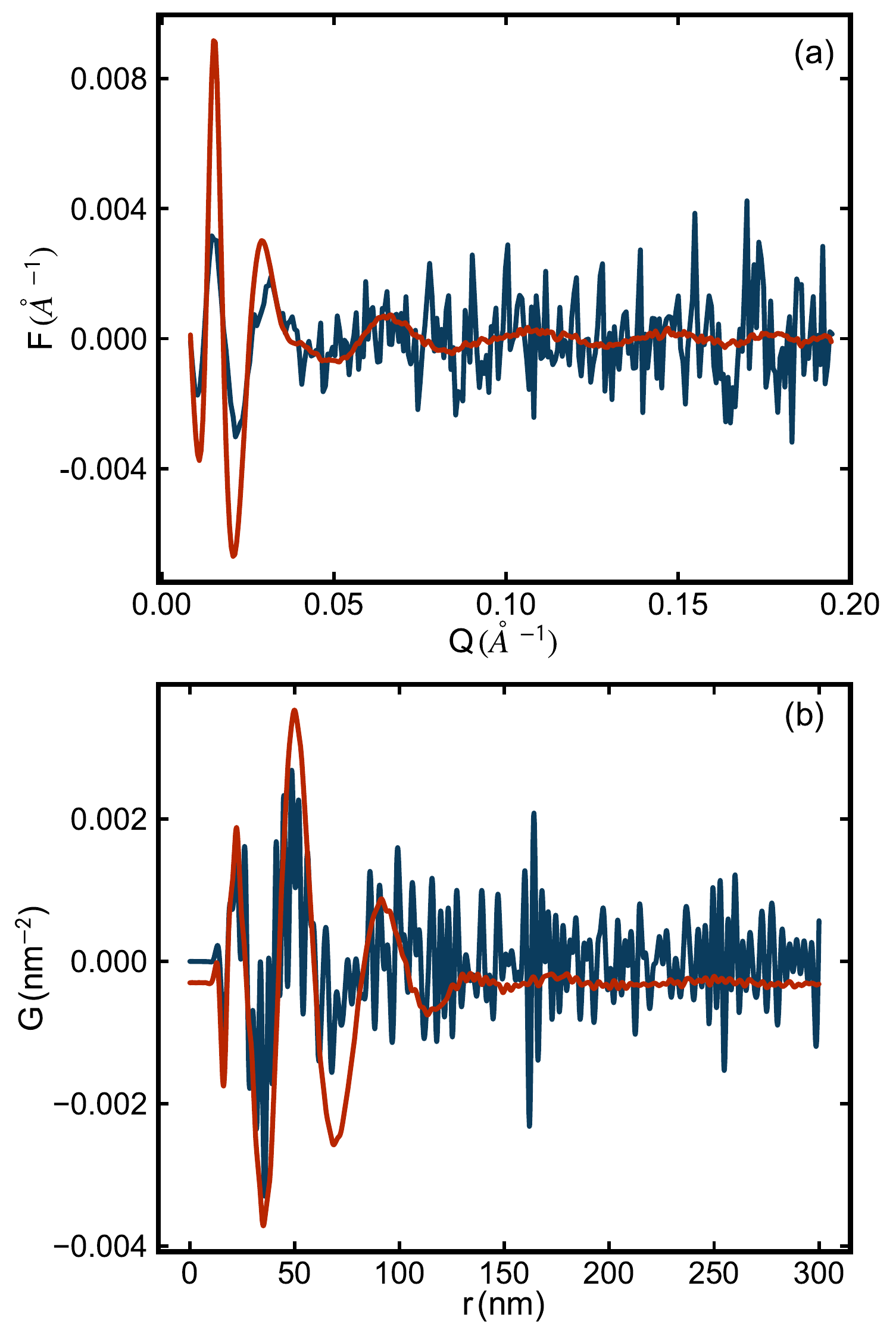}
    \label{fig:inhouse_vs_synchrotron}
    \caption{(a) Reduced structure functions $F(Q)$ and (b) PDFs $G(r)$ of the SiO$_2$ NPA sample.
    Blue is from data collected at Columbia University using a SAXSLAB
    (Amherst, MA) instrument with a 2-hour (7200~s) scan exposure time for both $I(Q)$ and $P(Q)$ measurements.
    Red is from data collected at beamline 11-BM, NSLS-II with 30~s scan exposure time for both $I_m(Q)$ and $P(Q)$ measurements.}
\end{figure}

\begin{figure}
    \includegraphics[width=0.58\columnwidth]{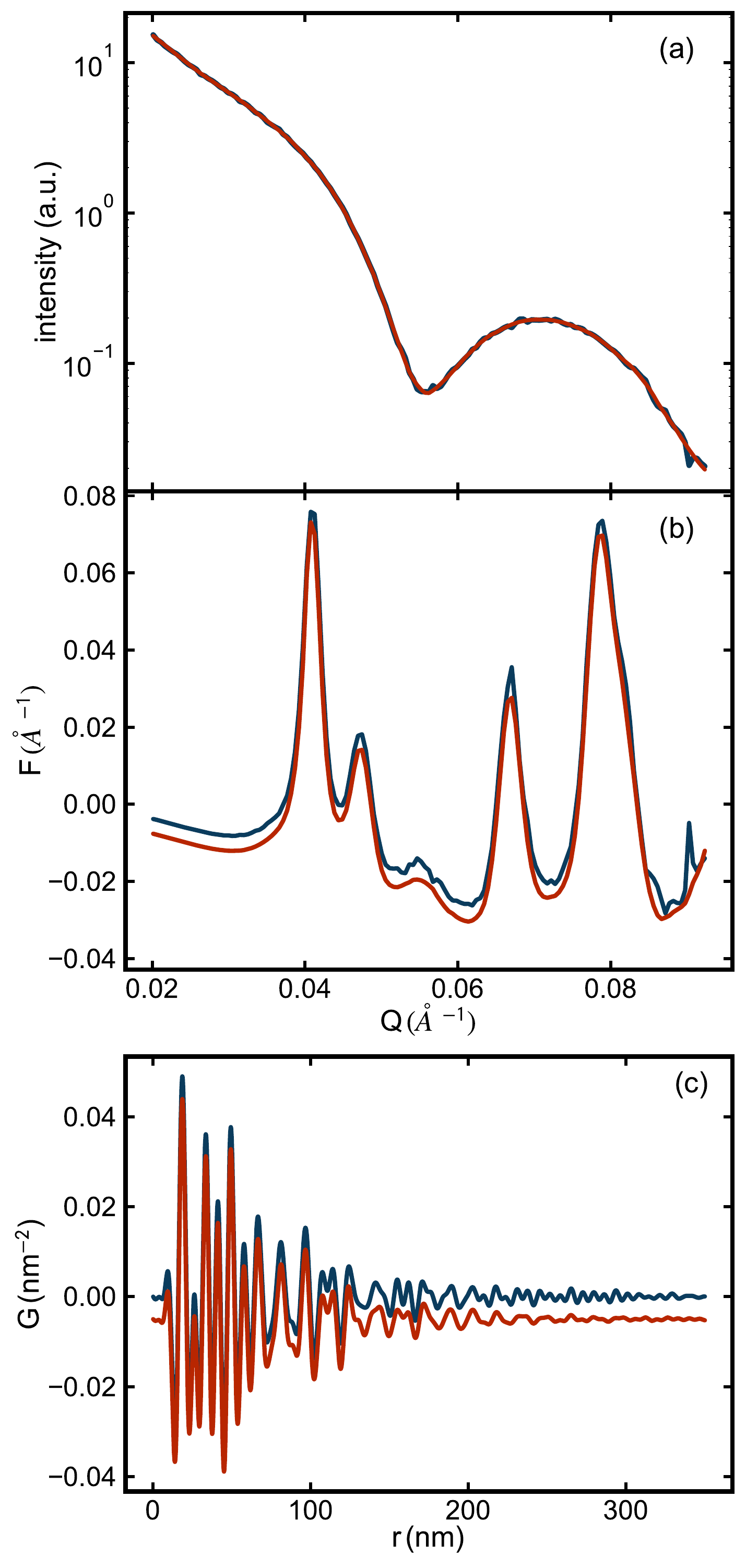}
    \label{fig:smoothed_Cu2S_form_factor}
    \caption{(a) Form factor signal from Cu$_2$S NPs. Blue is the raw data collected at an in-house instrument and red is the data smoothed by applying a Savitzky-Golay filter with window size 13 and fitted polymer order 2. (b) reduced structure functions, $F(Q)$, and (c) PDFs, $G(r)$ from the Cu$_2$S NPA sample. In both panel, blue represents the data processed with raw form factor signal and red represents the data processed with smoothed form factor signal. Curves are offset from each other slightly for ease of view.}
\end{figure}

In this section, important effects related to the data quality are illustrated.
In general, for a successful \saspdf experiment, it is crucial to achieve a high signal-to-noise ratio throughout the entire $Q$-range for both the form factor and sample measurements.
Figs.~\ref{fig:spot_exposure_example} and~\ref{fig:ff_spot_exposure_example} show the effect of insufficient counting statistics in the sample and form factor measurements, respectively.
Fig.~\ref{fig:inhouse_vs_synchrotron} compares the data quality from an in-house instrument and a synchrotron source.
Finally, Fig.~\ref{fig:smoothed_Cu2S_form_factor} shows the remedial effect of smoothing data from in-house measured form factor with insufficient statistics.
The proper remedy is to measure with sufficient statistics in the first place.
\begin{figure}
    \includegraphics[width=0.8\columnwidth]{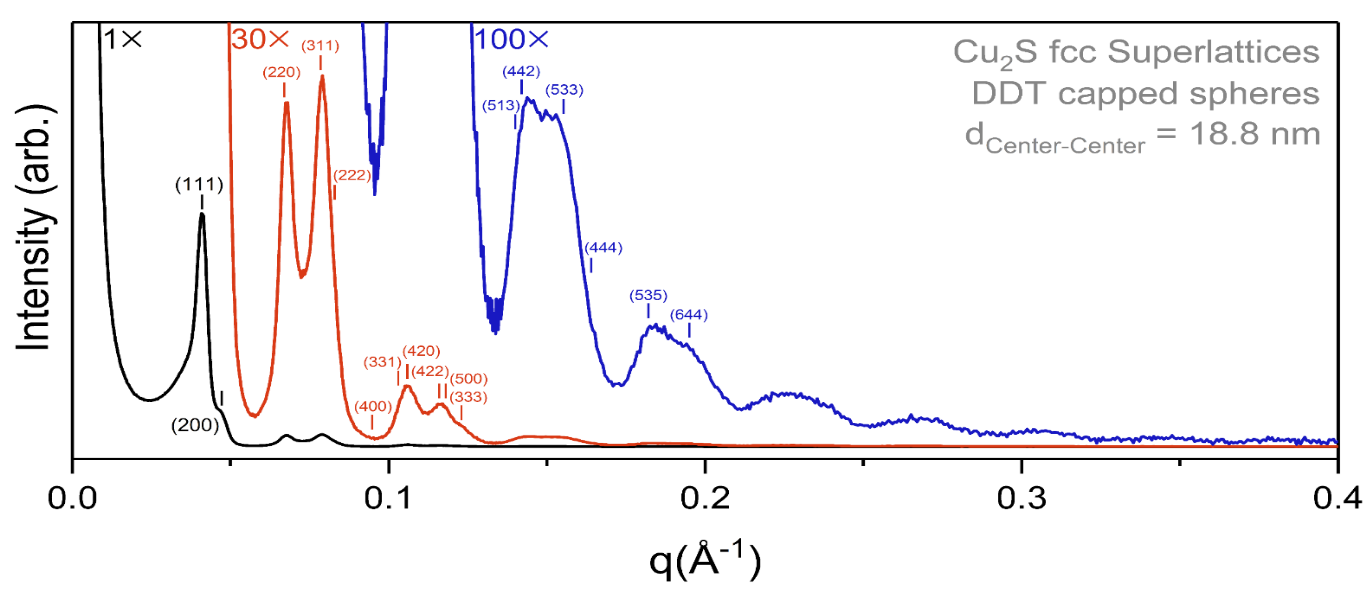}
    \label{fig:Cu2S_inhouse_Iq_indexing}
    \caption{Semi-quantitative structural analysis on Cu$_2$S NPA sample.}
\end{figure}

\bibliography{18cl_superlatticeAssembly}
\bibliographystyle{iucr}

\end{document}